\documentclass[fleqn,twoside]{article}
\usepackage{espcrc2}
\usepackage{amsmath}
\usepackage{amssymb}

\usepackage{graphicx}
\usepackage{multirow}
\usepackage{mycommands-short}

\title{Hidden and open charm at Belle (and elsewhere)}

\author{B.D.~Yabsley\address{High Energy Physics Group, 
				School of Physics, A28          \\
				University of Sydney. NSW 2006, 
				Australia}}

\begin{document}

\begin{abstract}
	There have been great advances in charm physics at the \bmes-factories in
	recent years, including the observation of \dz-\dzbar\ mixing and the
	discovery of many new hidden-charm states, some of them presumably
	exotic in structure. This talk reviews some of the recent Belle results,
	with emphasis on the hidden-charm sector.
	Evidence for two new charged states is presented for the first time at this conference.
\end{abstract}

\maketitle

\section{INTRODUCTION}

It has been a pleasure to present this review at BEACH 2008,
not only because of the organisers' invitation and the quality of the conference,
but also because I never made it to BEACH 2006!\footnote{For those who haven't heard the story,
	I wound up in hospital for an extended stay, in the weeks leading up to the conference.
	As for the food this year in Columbia SC: 
	as well as being excellent, it seems to have been bug-free.}
I was to survey charm mixing at that meeting,
right on the threshold of the beautiful results of 2007.
To make up for that missed talk,
I open this one by looking back at mixing over the last two years
(Section~\ref{section-mixing}). The remainder is given over to the hidden-charm sector,
which has been such an unexpected source of discoveries at the \bmes-factories. 
Summarizing results on
the $X(3872)$ (Section~\ref{section-x3872}),
the $\dmes^{(*)}\dbar{}^{(*)}$ states (Section~\ref{section-x3940}),
$Y(3940)$ (Section~\ref{section-y3940}),
initial state radiation (Section~\ref{section-isr}),
and the sector as a whole (Section~\ref{section-xyz-summary}), 
we conclude with candidates for \emph{charged} states with hidden charm,
including new results first presented at this conference (Section~\ref{section-xyz-z1-z2}).
If confirmed, of course, such states are manifestly exotic in structure.


\section{Charm mixing}
\label{section-mixing}

Since the summer of 2006 there has been a procession of important mixing results:
\begin{itemize}
  \item	Belle's \ycp\ measurement in $\dz\to \kplus\kmin$ and $\pi^+\pi^-$~\cite{belle-ycp},
	providing robust evidence for mixing:
	a strongly data-driven resolution function is used;
	successive running periods are taken into account,
	with the \dz\ lifetime recovered in each;
	and only binned fits are used, with good fit quality throughout.
  \item	The time-dependent $\dz\to\ks\pi\pi$ Dalitz analysis~\cite{belle-kspipi},
	following CLEO's method~\cite{cleo-kspipi} but with sixty times their data sample,
	giving direct access to the mixing parameters $(x,y)$
	at three-permille precision.
	The result only constitutes weak evidence for mixing in isolation ($2.2\sigma$), 
	but plays an important role in the ``average''
	over all measurements: it anchors the allowed region in the $(x,y)$ plane,
	as there is no rotation of the parameters by ill-constrained strong phases.
	A lifetime difference $y$ on the low side of the \ycp\ result is favoured.
  \item	A comprehensive update~\cite{belle-semilep}
	of the $\dz\to\kmes^{(*)+}\ell^-\anu$ analysis,
	with an expanded dataset and the addition of muonic modes.
	As expected, this is not competitive with the evidence 
	in other channels, but the limit on the mixing rate
	($R_M < 6.1 \times 10^{-4}$ at 90\% confidence) is tight enough to exclude
	regions of parameter space allowed by older hadronic results:
	\emph{e.g.}\ FOCUS' \ycp~\cite{focus-ycp} would correspond to 
	$R_M \sim 5.9 \times 10^{-4}$.
\end{itemize}
\begin{figure}
  \includegraphics[width=6cm]{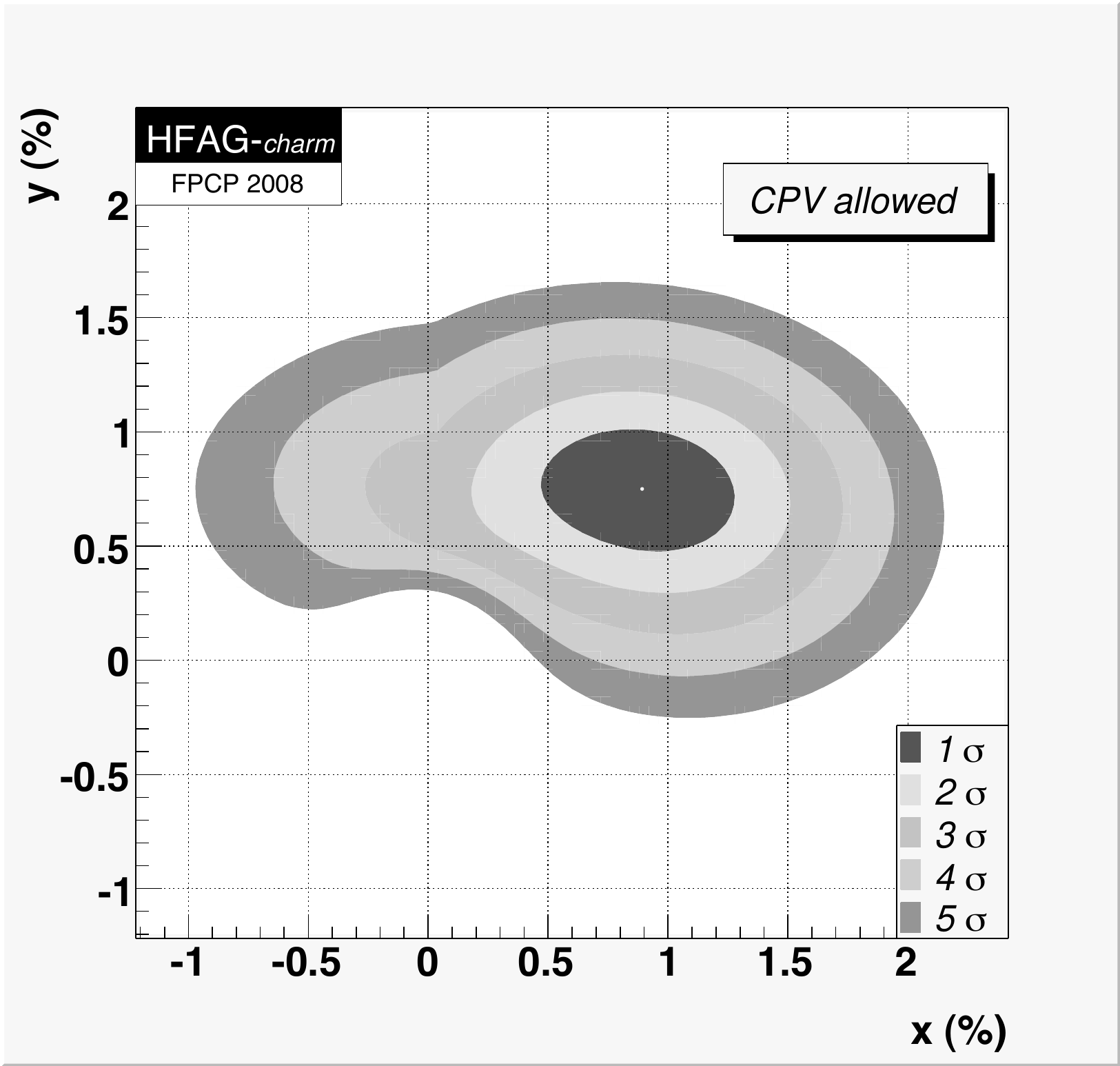}
  \caption{Combined results for the charm mixing parameters $(x,y)$,
	from HFAG~\protect\cite{hfag-charm}.}
  \label{fig-hfag-mixing}
\end{figure}
The combined (``world-average'') constraints on $(x,y)$
from the Heavy Flavor Averaging Group~\cite{hfag-charm}
are shown in Fig.~\ref{fig-hfag-mixing}.
Among the other results that contribute to the average, the most important
are those in $\dz\to\kplus\pi^-$, where BaBar found evidence for mixing 
at the $3.9\sigma$ level~\cite{babar-kpi}.
An earlier Belle measurement~\cite{belle-kpi}, with similar sensitivity,
found weaker
evidence because of the position of the central value.
CDF have since published a compatible $3.8\sigma$ result
in the same channel~\cite{cdf-kpi}.

The average is driven by $\ks\pi^+\pi^-$ in $x$, 
and by this and \ycp\ in $y$;
the $\kplus\pi^-$ mode contributes to 
the exclusion of $(x,y) \approx (0,0)$,
producing the ``bite'' out of the confidence regions at the lower left.
(BaBar's $\dz\to\kplus\pi^-\pi^0$ and $\kplus\pi^-\pi^+\pi^-$ results,
presented by V.~Santoro at this meeting, also provide support.)
And by combining the $y$ average with rate measurements at their coherent  
$(\ket{\dz}\ket{\dzbar}-\ket{\dzbar}\ket{\dz})$ source,
CLEO-c have been able to place a serious constraint on the strong-phase 
difference $\delta_{\kmes\pi}$ for the first time~\cite{cleo-delta}.

CP violation parameters are not yet strongly constrained~\cite{hfag-charm},
although with mixing parameters $(x,y) \neq (0,0)$ at high confidence,
large CP violation (\emph{i.e.}\ due to physics outside the Standard Model)
is now within reach.
For example, the decay rate asymmetry between \dz\ and \dzbar\ decaying to 
$\kplus\kmin$ (and $\pi^+\pi^-$) is already constrained to
$A_\Gamma = (0.123\pm 0.248)\%$~\cite{hfag-charm}, based on side-measurements
in the Belle~\cite{belle-ycp} and BaBar~\cite{babar-ycp} \ycp\ analyses.
Time-integrated CP asymmetries for $\dz \to \kplus\kmin$ and $\pi^+\pi^-$
have also been measured by both collaborations~\cite{babar-acp,belle-acp}.


\section{The $X(3872)$}
\label{section-x3872}

The first of the anomalous charmonium-like states to be seen,
the $X(3872)$ is the best-understood, although its structure is still in dispute.
We first review the state of knowledge at Beauty 2006~\cite{beauty2006-yabsley}
(shortly after the previous BEACH) before discussing recent results.

\subsection{The state of play in 2006}

The $X(3872)$ is narrow ($\Gamma< 2.3\,\mev$~\cite{x3872-belle-discovery}),
with a prominent $\pi^+\pi^-\psi$ decay~\cite{x3872-belle-discovery,x3872-cdf-confirmation,x3872-d0-confirmation,x3872-babar-confirmation}:
$\br(X\to\pi^+\pi^-\psi) > 4.2\%$~\cite{x3872-babar-inclusive},
based on BaBar's inclusive analysis.
Its mass, $(3871.2 \pm 0.5)\,\mev$ ($S=1.4$)~\cite{pdg2006}, is $0.6\,\mev$ below
the $\dstar\dbar$ threshold~\cite{pdg2006,dmass-cleo}, although at only $1\sigma$
this point still needs to be firmly established.
The state seems to include \emph{some} compact component (presumably \ccbar),
because production in \prot\aprot\ collisions is $\psi'$-like:
only $(16\pm5\pm2)\%$ is due to $b$-hadron decays, the rest being prompt~\cite{x3872-bauer}.
It is not an isovector, as charged partners $X^\pm$ are not seen~\cite{x3872-babar-charged}.

The $X(3872)$ is even under charge conjugation, as $X\to\gamma\psi$
is seen~\cite{x3872-belle-radiative,x3872-babar-radiative},
and there is evidence for decay to $\pi^+\pi^-\pi^0 \psi$ 
(consistent with sub-threshold $\omega\psi$) at a comparable rate
to the discovery mode~\cite{x3872-belle-radiative}.
Isospin is thus badly broken in these decays.
Together with the coincidence in masses, this is the key evidence
for the \dstar\dbar\ molecular model, where isospin violation is natural
(($m_{\dstarp}+m_{\dmin}) - (m_{\dstarz}+m_{\dzbar}) \gg E_{\text{binding}}$).
There are further constraints on the the decay amplitude
and quantum numbers:
$X\to\pi^+\pi^-\psi$ is dominated by $X\to\rho\psi$, with $L=0,1$~\cite{x3872-cdf-dipion};
and $J^{PC} = 1^{++}$ or $2^{-+}$
(from~\cite{x3872-cdf-angular}, superseding~\cite{x3872-belle-angular}).

Unresolved issues at the time included the relative production rate
in \bplus\ \emph{vs} $\bz\to\kmes\,X$,
and possible mass-splitting in these two modes.
These quantities can discriminate
between molecular and diquark-antidiquark models: 
inconclusive results were available from BaBar~\cite{x3872-babar-xu-xd}.
Belle's peak in $\bmes\to\kmes\dz\dzbar\pi^0$,
consistent with a large $X\to\dz\dzbar\pi^0$ rate~\cite{x3872-belle-ddpi},
was still awaiting confirmation.
And a number of experimental loose ends could be discerned:
searches for $\pi^0\pi^0\psi$ (forbidden for a $C$-even $X$);
$\gamma\psi'$ (its rate an important diagnostic);
$\pi^+\pi^-\eta_c$ (prominent for a $2^{-+}$ charmonium);
and $\dmes\dbar\gamma$ (for its rate and lineshape).

\subsection{Recent results}

\begin{figure*}
  \includegraphics[width=6.5cm]{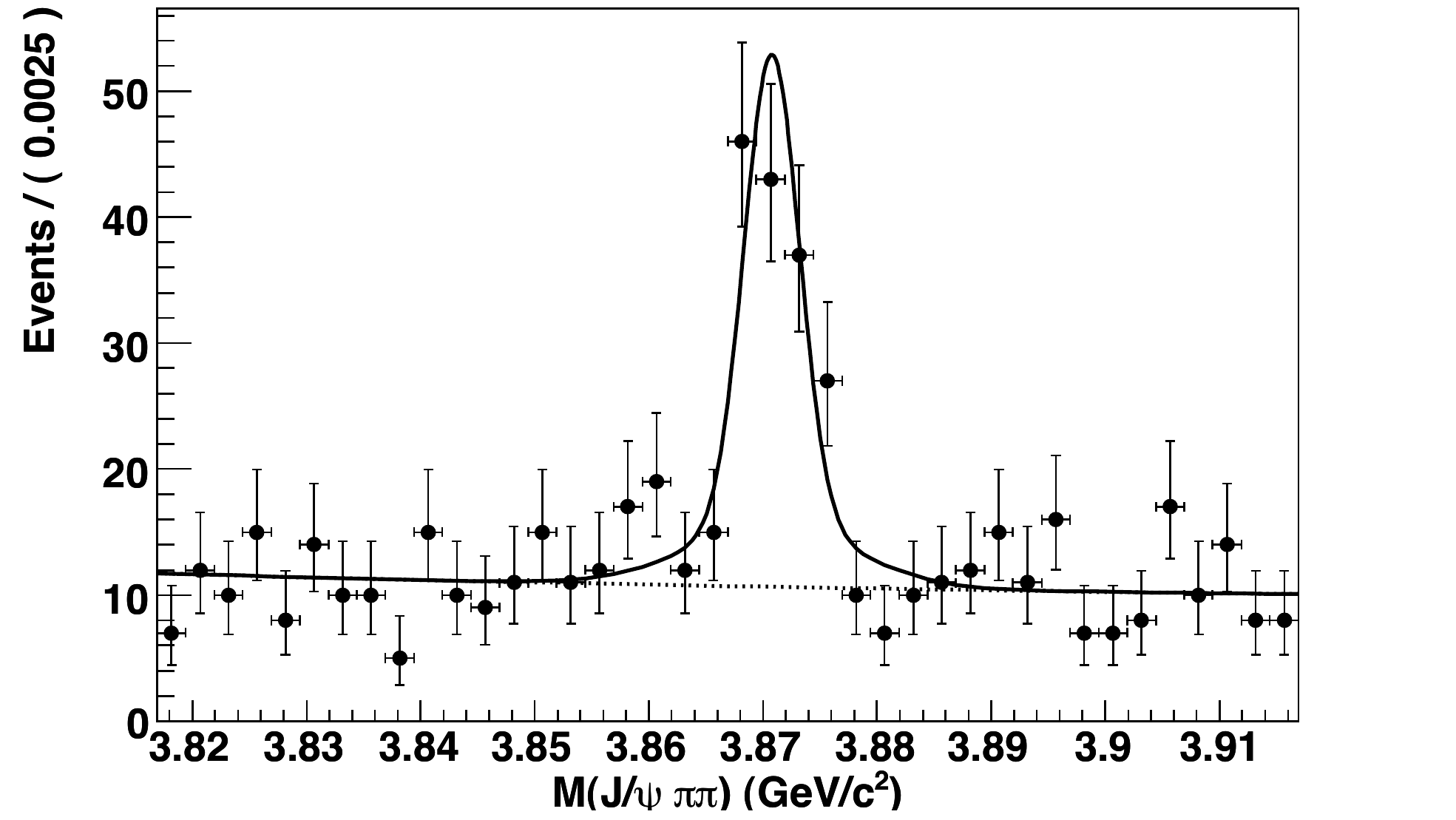}
  \includegraphics[width=6.5cm]{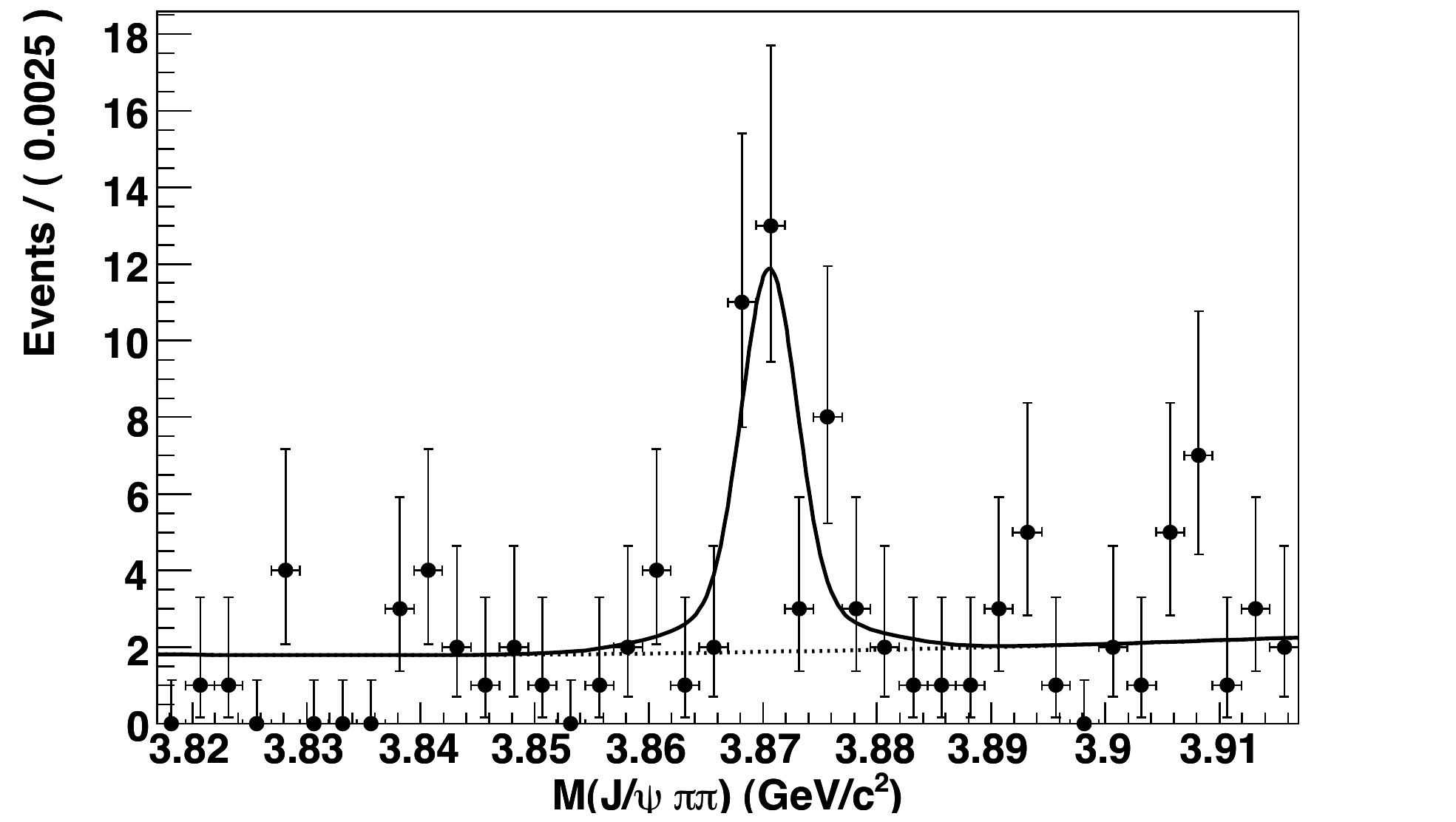}
  \caption{Distributions of $\pi^+\pi^-\psi$ invariant mass 
	at Belle~\protect\cite{x3872-belle-pipipsi-update},
	with fits for $X(3872)$ signal shown, for
	(upper) $\bplus\to\kplus\pi^+\pi^-\psi$, and 
	(lower) $\bz\to\ks\pi^+\pi^-\psi$ events.}
  \label{fig-belle-pipipsi-update}
\end{figure*}

These loose ends have not yet been addressed,\footnote{After the BEACH conference,
	BaBar presented evidence for the decay $X\to\gamma\psi'$ 
	at Philadelphia~\protect\cite{x3872-babar-gamma-psiprime}.}
but Belle has revisited the discovery mode
with a $605\,\ifb$ sample ($657 \times 10^6\;\BBbar$)~\cite{x3872-belle-pipipsi-update}.
$X\to\pi^+\pi^-\psi$ mass peaks are shown in 
Fig.~\ref{fig-belle-pipipsi-update},
for both $\bplus\to\kplus\,X$ and
$\bz\to\ks\,X$ (at $6.5\sigma$, the first formal observation of this mode).
Belle finds comparable branchings,
with $R \equiv \br_{\kz X}/\br_{\kplus X} = 0.94 \pm 0.24 \pm 0.10$,
in some tension with BaBar's
$R = 0.41 \pm 0.24 \pm 0.05$~\cite{x3872-babar-pipipsi-update}.
It is important not to over-interpret these results:
the claim that the molecular model
predicts $R \lesssim 0.1$~\cite{x3872-braaten-kusunoki}
has been withdrawn by one of its authors~\cite{x3872-braaten-lu}.
The mass splitting is more straightforward:
Belle finds $\delta m \equiv M_{\kplus X} - M_{\ks X} = (+0.22\pm 0.90 \pm 0.27)\,\mev$
(\emph{cf.}\ BaBar's $(+2.7 \pm 1.6 \pm 0.4)\,\mev)$,
providing no evidence for the diquark-antidiquark model,
although it cannot rule it out.

It is now established that the $X(3872)$ decays to open charm,
with BaBar confirming $X\to\dz\dzbar\pi^0$ with large
branching fraction~\cite{x3872-babar-ddpi}.
The mechanism is still unknown:
is it $X \to \dstarz\dzbar$,
or more properly $\dz\dzbar\pi^0$ (and $\dz\dzbar\gamma$)
with some nontrivial lineshape that probes the structure of the $X(3872)$?
Current analyses do not speak to this. 
BaBar sees a $(3875.1^{+0.7}_{-0.5} \pm 0.5)\,\mev$ mass peak
(\emph{cf.}\ Belle's $(3875.2 \pm 0.7^{+0.3}_{-1.6} \pm 0.8)\,\mev$),
well above that in the discovery mode,\footnote{Also at Philadelphia,
	Belle presented a new (preliminary) $X\to\dstar\dbar$ analysis
	with $M = (3872.6^{+0.5}_{-0.4} \pm 0.4)\,\mev$~\protect\cite{x3872-belle-ddpi-update}.}
although this is not straightforward to interpret.


\section{States above open-charm threshold in $\epem \to \psi^{(\prime)} \dmes^{(*)} \dbar{}^{(*)}$
}
\label{section-x3940}

Belle has also updated its results on states decaying to open charm,
$\dmes^{(*)} \dbar{}^{(*)}$,
using a refinement of the recoil-mass method~\cite{x3940-belle-refined}.
After reconstruction and mass-constraint of $\psi\to\ell^+\ell^-$,
a \dz, \dplus, or \dstarp\ is also selected and refitted to its nominal mass;
the remaining unreconstructed meson is then tagged, requiring
$|\mrecoil(\psi\dmes^{(*)})-m_{\text{tag}}|<70\,\mev$,
about twice the resolution on this quantity.
$\mrecoil(\psi\dmes^{(*)})$ is then constrained
to $m_{\text{tag}}=m_{\dmes^{(*)}}$:
the resolution on $M(\dmes^{(*)}\dbar{}^{(*)})$
improves by a factor of 3--10.
Resulting spectra are shown in Figure~\ref{fig-x3940-belle-refined}.
Simultaneous fits are performed to the tagging meson
mass signal and sideband regions (the latter to model the background),
including cross-feed, 
threshold functions to represent non-resonant contributions,
and relativistic S-wave Breit-Wigner signal terms.

\begin{figure}
  \includegraphics[width=6.0cm]{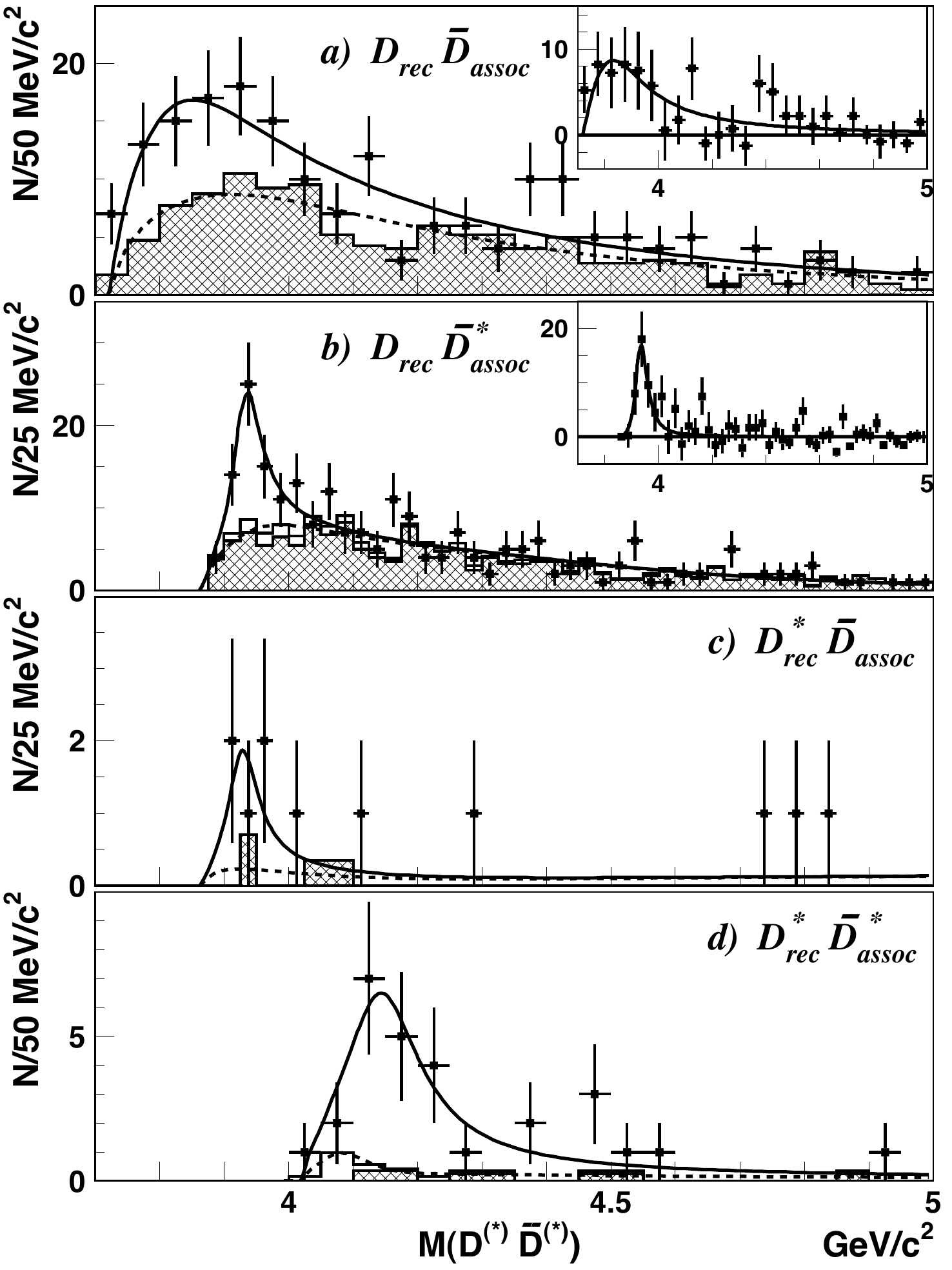}
  \caption{Invariant mass of $\dmes^{(*)} \dbar{}^{(*)}$ 
	in $\epem \to \psi \dmes^{(*)} \dbar{}^{(*)}$ events
	at Belle~\protect\cite{x3940-belle-refined}.
	Solid curves show fits to signal, 
	background (dashed) modelled using mass sidebands (shaded),
	cross-feed, and nonresonant terms.
	In the upper plots, background-subtracted distributions are
	shown inset.}
  \label{fig-x3940-belle-refined}
  \vspace*{5ex}

  \includegraphics[width=5.5cm]{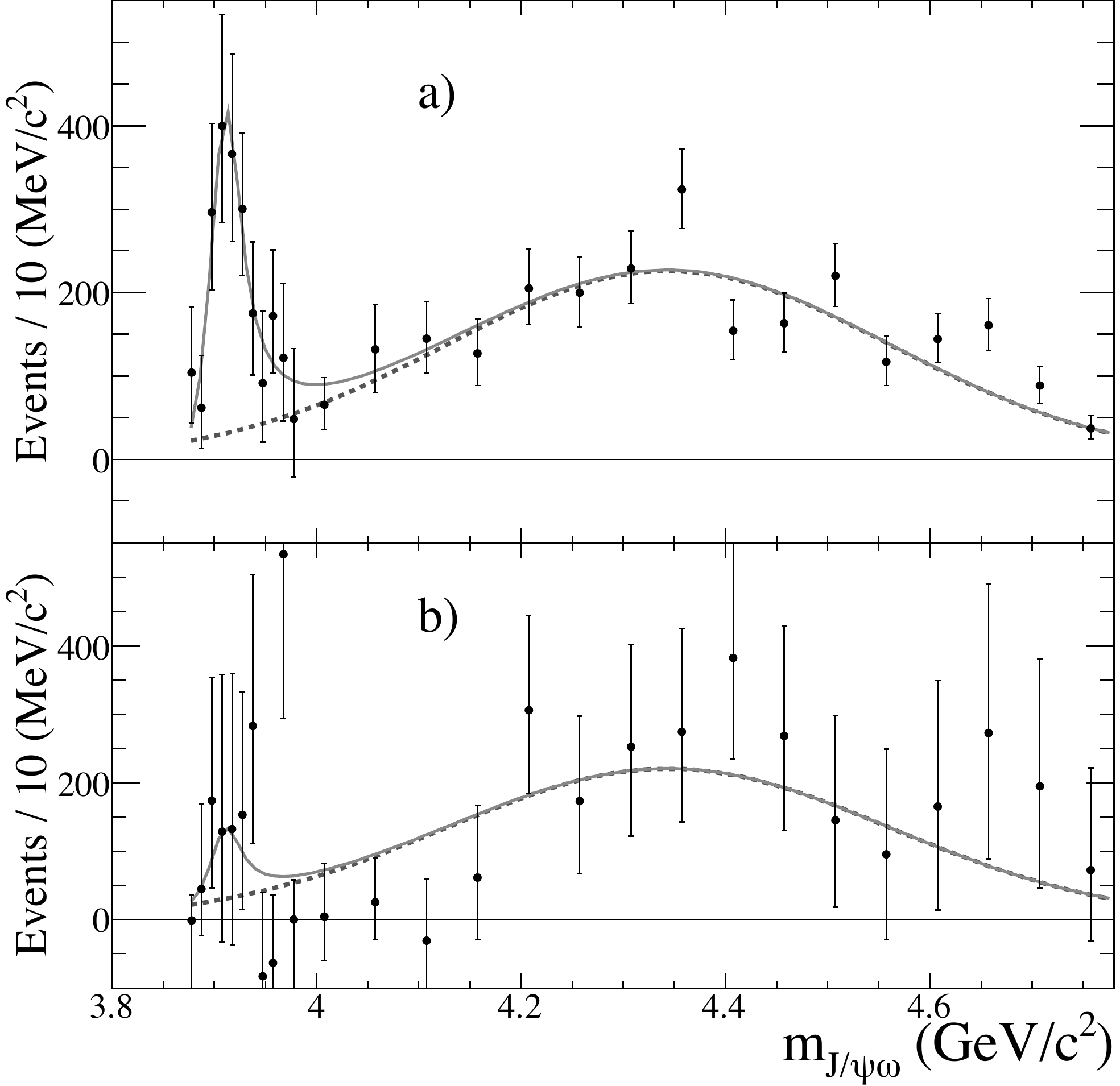}
  \caption{BaBar's confirmation~\protect\cite{y3940-babar}
	of the $Y(3940)$ at $\omega\psi$ threshold in
	(a) \bplus\ and (b) $\bz\to\kmes\omega\psi$.
	The fits are discussed in the text.}
  \label{fig-y3940-babar}
\end{figure}

The $X(3940) \to \dmes\dbar{}^*$ signal is confirmed (at $6\sigma$),
with $\sigma(\epem\to\psi\,X) \times \br(X\to\dmes\dbar{}^*) = 
(13.9^{+6.4}_{-4.1} \pm 2.2)\,\fb$.
A clear resonant peak ($5.5\sigma$) is also seen in $\dstar\dbar{}^*$,
called $X(4160)$ by Belle, with $\sigma\times\br = (24.7^{+12.8}_{-8.3} \pm 5.0)\,\fb$.
(A broad enhancement, difficult to fit, is also seen at $\dmes\dbar$ threshold.) 
These effective cross-sections are comparable to those of the other
$\epem\to\psi(nS)\,\eta_c(mS)$ modes ($m,n \in \{1,2\}$) seen by Belle,
all of which are close to $20\,\fb$.
These results are thus consistent with the $X(3940,4160)$ being \ccbar\ states,
although this underlines the fact that the double charmonium production
mechanism is not understood.


\section{The\boldmath\ $Y(3940)$ at $\omega\psi$ threshold, \& BaBar}
\label{section-y3940}

In 2005 Belle reported a significant enhancement ``$Y(3940)$''
at $\omega\psi$ threshold, in a $\bmes\to\kmes\omega\psi$ sample~\cite{y3940-belle}.
This is (probably) distinct from the $X(3940)$, an $\eta_c(3S)$ candidate
seen in $\epem\to\psi\,X$ (discussed in the previous section);
and the $Z(3930)$, likely the $\chi_{c2}(2P)$,
seen in $\gamma\gamma\to\dmes\dbar$~\cite{z3930-belle}.
Unlike those states, the $Y(3940)$ is difficult to interpret
as charmonium---due to the prominent $\omega\psi$ mode, for example---and
hard to understand even as an exotic, such as a $\ccbar g$ hybrid.
It has thus been the least-believed of the new charmonium-like,
so-called ``XYZ'' states.

In November, a spectacular confirmation of the $Y(3940)$ was announced
by BaBar~\cite{y3940-babar}: invariant mass plots, based on fits to the
\bmes-meson yield in each bin, are shown in Fig.~\ref{fig-y3940-babar}.
Mass-dependent corrections for resolution and efficiency effects are applied,
and the $Y$ yield is fitted to the \bplus\ and \bz\ samples simultaneously,
with the branching ratio floated. 
Gaussian background and S-wave Breit-Wigner signal terms are used
(\emph{cf.}\ Belle's use of a threshold function $\propto q^*(M)$
for the background).
The signal term is found to have $M = (3914.6^{+3.8}_{-3.4} \pm 1.9)\,\mev$
and $\Gamma = (34^{+12}_{-8} \pm 5)\,\mev$, both smaller than
Belle's values.\footnote{See also the discussion on this point in the Appendix.}
Decays from the charged \bmes-meson are favoured:
$R_Y = \br_{\bz}/\br_{\bplus} = 0.27^{+0.28}_{-0.23} {}^{+0.04}_{-0.01}$,
to be compared with
$R_{\text{non-res}} = 0.97^{+0.23}_{-0.22} {}^{+0.03}_{-0.02}$
for $\bmes\to\kmes\omega\psi$ events away from the $Y(3940)$ mass peak.


\section{Vector states in ISR (following BaBar)}
\label{section-isr}

The unexpected (and presumably exotic) peak
$Y(4260)\to\pi^+\pi^-\psi$ seen by BaBar~\cite{y4260-babar-discovery}
in initial state radiation (ISR) events,
and confirmed by CLEO~\cite{y4260-cleo-scan,y4260-cleo-isr},
has also been seen by Belle~\cite{y4260-belle-prl}.
Furthermore, the cross-section around $4050\,\mev$
appears to be nontrivial in the Belle data;
fits with interfering Breit-Wigner terms find a significant enhancement
there, but with large systematic errors as fit conditions are varied.

The broad peak seen by BaBar in $\pi^+\pi^-\psi'$~\cite{y4360-babar}
is also confirmed by Belle~\cite{y4360-y4660-belle},
but split into two distinct peaks at
$(4361\pm 9\pm 9)$ and $(4664\pm 11\pm 5)\,\mev$.
The two sets of results appear consistent with a two-peak
structure, and the effect of the larger sample (and better luck) of Belle.

None of these enhancements are seen in ISR production of $\dmes^{(*)}\dbar{}^{(*)}$:
for example, BaBar set a limit on $Y(4260)$ decays to this final state of
$\br_{\dmes\dbar}/\br_{\pi^+\pi^-\psi} < 1.0$ at 90\% confidence~\cite{isr-ddbar-babar}.


\section{Hidden-charm states: $X$, $Y$, \ldots\ and $Z$}
\label{section-xyz}

\label{section-xyz-summary}

The contours of the hidden-charm sector are increasingly clear,
and if a comprehensive theoretical account is still elusive,
we at least know some of the features it must have.

The \ccbar\ bound states expected below \dmes\dbar\ threshold have all been seen.
Above threshold,
the $Z(3930)\to\dmes\dbar$ is widely believed to be the $\chi_{c2}(2P)$;
the $X(3940)\to\dmes\dbar{}^*$ is plausibly the $\eta_c(3S)$;
and the new $X(4160)\to\dstar\dbar{}^*$ may also be a radial excitation,
\emph{e.g.}\ $\eta_c(3S)$ or $\chi_{c0}(3P)$~\cite{x4160-chao}.
The $Y(3940)\to\omega\psi$ has been confirmed, but what is it:
the $\chi_{c1}(2P)$ at the ``wrong'' mass,
decaying via rescattering~\cite{eichten-quarkonia-transitions}?
A $\ccbar g$ hybrid? (The mass likewise seems wrong for this.)
Some other kind of exotic, or a nontrivial effect of the threshold?

We know at least that effects due to dynamics, opaque to many experimentalists,
are important. The $X(3872)$ is \emph{right at} $\dz\dbar{}^{*0}$ threshold:
on general grounds, $\ket{\dz}\ket{\dbar{}^{*0}}$ is involved \emph{somehow}
in the structure of the state.
The molecular model is still popular, despite occasional reports of its demise;
we note that Thomas and Close have recently tried to impose some order on the
literature on pion exchange and other sources of binding~\cite{thomas-close-molecular}.
The $\pi^+\pi^-\psi^{(\prime)}$ peaks seen in ISR---the $Y(4260)$ and its friends---may
be another case in point: any interpretation must account for multiple peaks
with varying $\psi$-\emph{vs}-$\psi'$ decay preferences.

There is a consensus that the phenomena imply
some non-\ccbar\ structure(s), but two caveats must be stressed:
not every peak is necessarily a state
(as discussed at Charm 2007~\cite{charm2007-panel}),
and the spectrum implied by any new structure must not be too rich.
For we do \emph{not} see a forest of new states.
In particular charged partners, which are both manifestly exotic 
and expected in many models, have been elusive.
Candidate charged states \emph{have} emerged in the last year, 
but they appear to be new cases, rather than partners of known states.

\subsection{The $Z(4430)^+\to\pi^+\psi'$ candidate state}
\label{section-xyz-z4430}

This first good candidate for a charged hidden-charm state conforms
to the pattern for the $X(3872)$ and $Y(4260)$:
hadronic decay to charmonium, far from threshold in the final state, 
with a clear peak above a (relatively) featureless environment.
Belle's study~\cite{z4430-belle} is straightforward,
forming a $\bmes\to\kmes\pi^\pm\psi'$ sample 
($\psi' \to \ell^+\ell^-$ and $\pi^+\pi^-\psi$)
with standard \bmes-reconstruction
and \qqbar\ continuum suppression techniques, 
finding a signal of over 2500 events with $\approx 90\%$ purity.
The Dalitz plot $(M^2(\kmes\pi^+),\,M^2(\pi^+\psi'))$
shows clear bands due to $\kstar\psi'$ and $\kmes_2^*(1430)\psi'$,
plus structure at $M^2(\pi^+\psi') \simeq 20\,\gev$.
A $\kstar(892)$ and $\kmes_2^*(1430)$ veto leads to the 
distribution shown in Fig.~\ref{fig-z4430-belle},
where a prominent $\pi^+\psi'$ peak can be seen.

\begin{figure}
  \includegraphics[width=5.5cm]{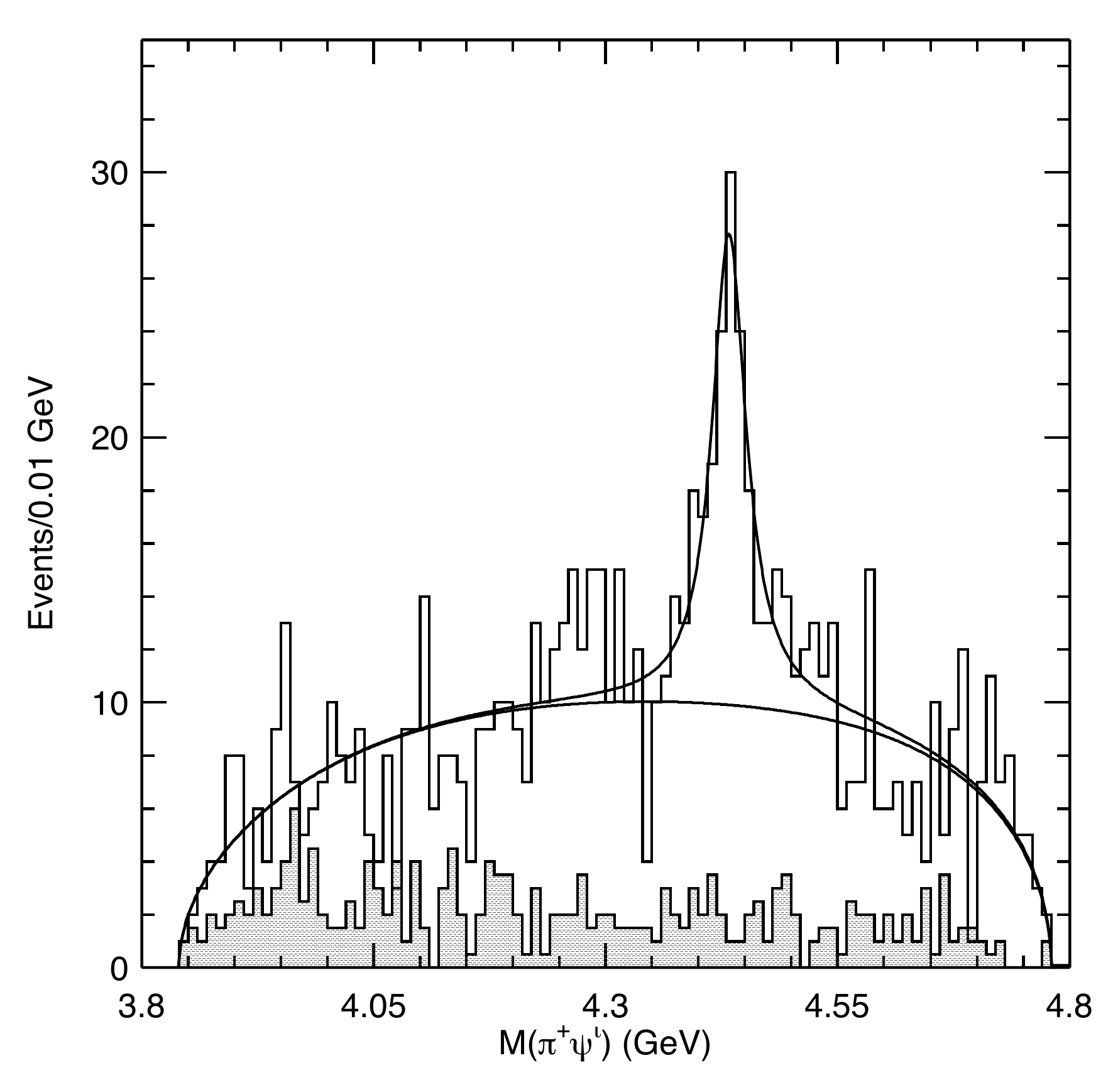}
  \caption{Invariant mass of $\pi^+\psi'$ in $\bmes\to\kmes\pi^\pm\psi'$ events
	at Belle~\protect\cite{z4430-belle};
	the $Z(4430)^+$ signal fit shown is described in the text.
	Data from the sideband in $\Delta E \equiv E_{\bmes} - E_{\text{beam}}$
	is shown shaded.}
  \label{fig-z4430-belle}
\end{figure}

The distribution is fitted with an S-wave Breit-Wigner term
over a double-threshold background function 
$q^*(Q^{1/2} + A_1 Q^{3/2} + A_2 Q^{5/2})$, where
$q^*$ is the $\pi^+$ momentum in the $\pi^+\psi'$ frame, and
$Q = M_{\text{max}} - M(\pi^+\psi')$.
The fit is good, with $\chi^2/n_{dof}=80.2/94$, 
and finds a $6.5\sigma$ signal peak.
Fits to various subsamples all find signals, with consistent masses;
the sole discrepancy is a disagreement in width
between the $\psi'\to\pi\pi\psi$ and $\ell^+\ell^-$ modes.
Possible interfering S-, P-, and D-wave $\kmes\pi$ terms are studied:
no such effect can mimic the signal peak
without producing additional dramatic structure.
The peak thus appears to be a good candidate for a new state, 
with $M = (4433 \pm 4 \pm 2)\,\mev$
and $\Gamma = (45^{+18}_{-13} {}^{+30}_{-13})\,\mev$.

\subsection{New: $Z^+\to\pi^+\chi_{c1}$ in $\bzbar \to \kmin \pi^+\chi_{c1}$?}
\label{section-xyz-z1-z2}

\begin{figure}
  \includegraphics[width=5.0cm]{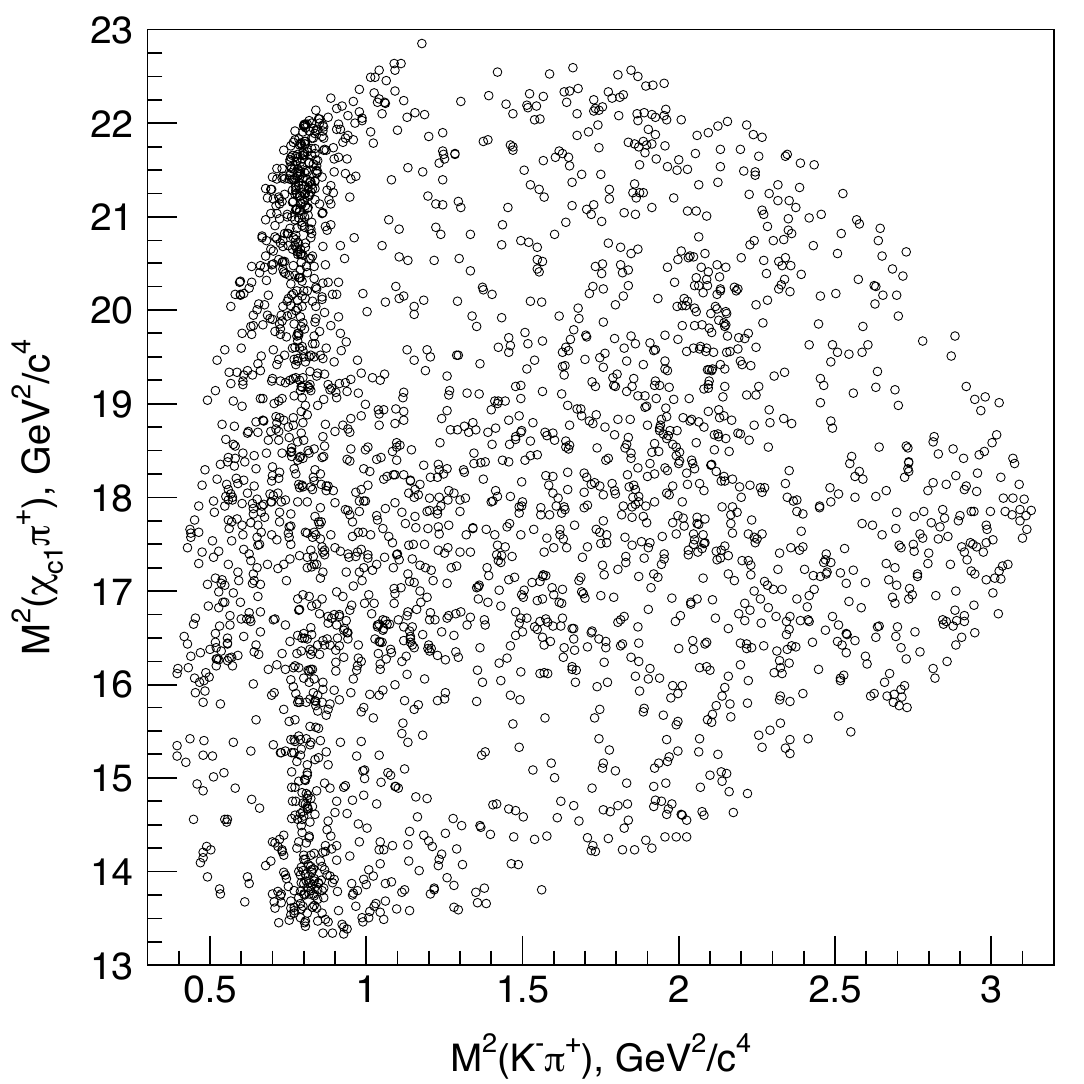}		\\[1.0ex]
  \includegraphics[width=5.5cm]{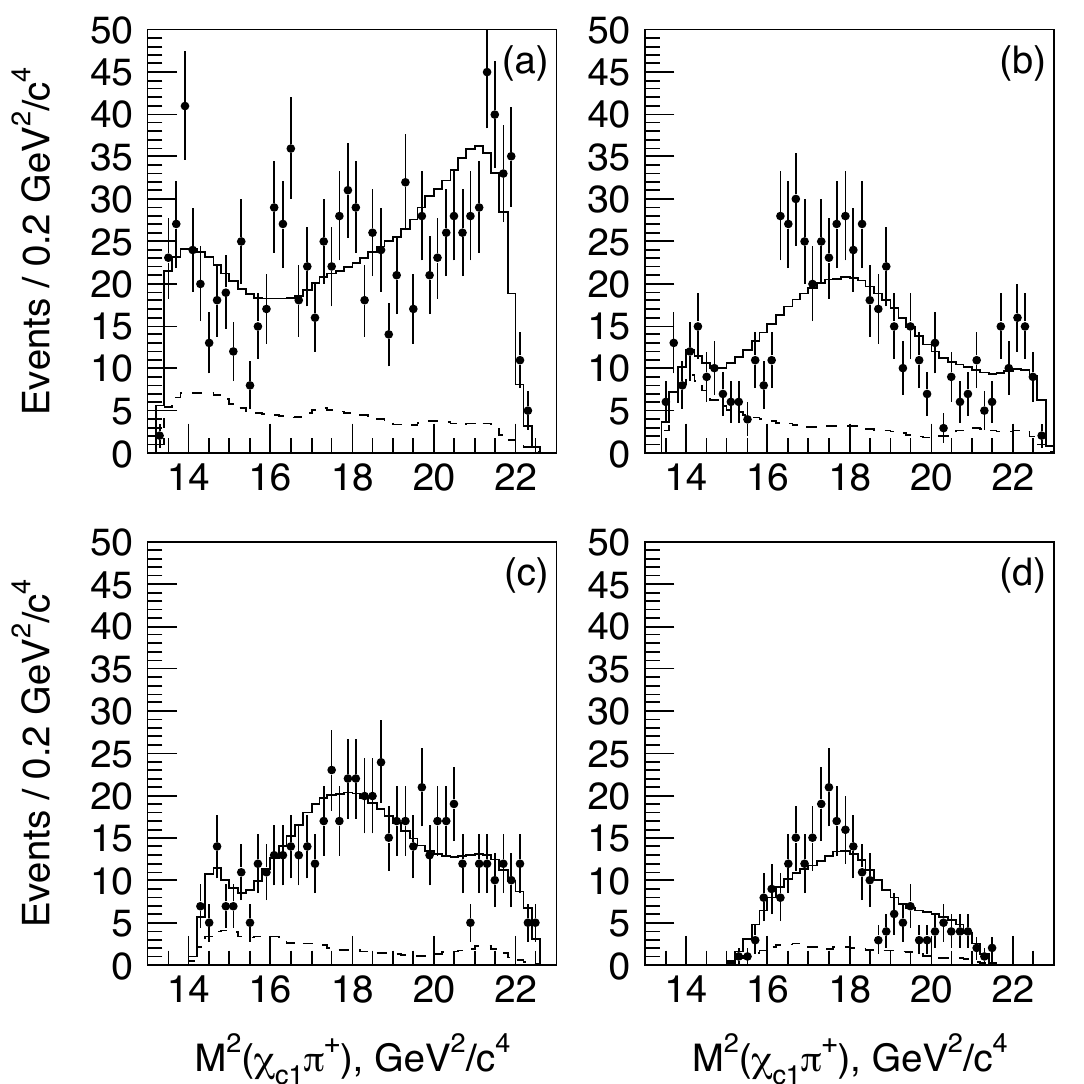}	\\[1.0ex]
  \includegraphics[width=5.5cm]{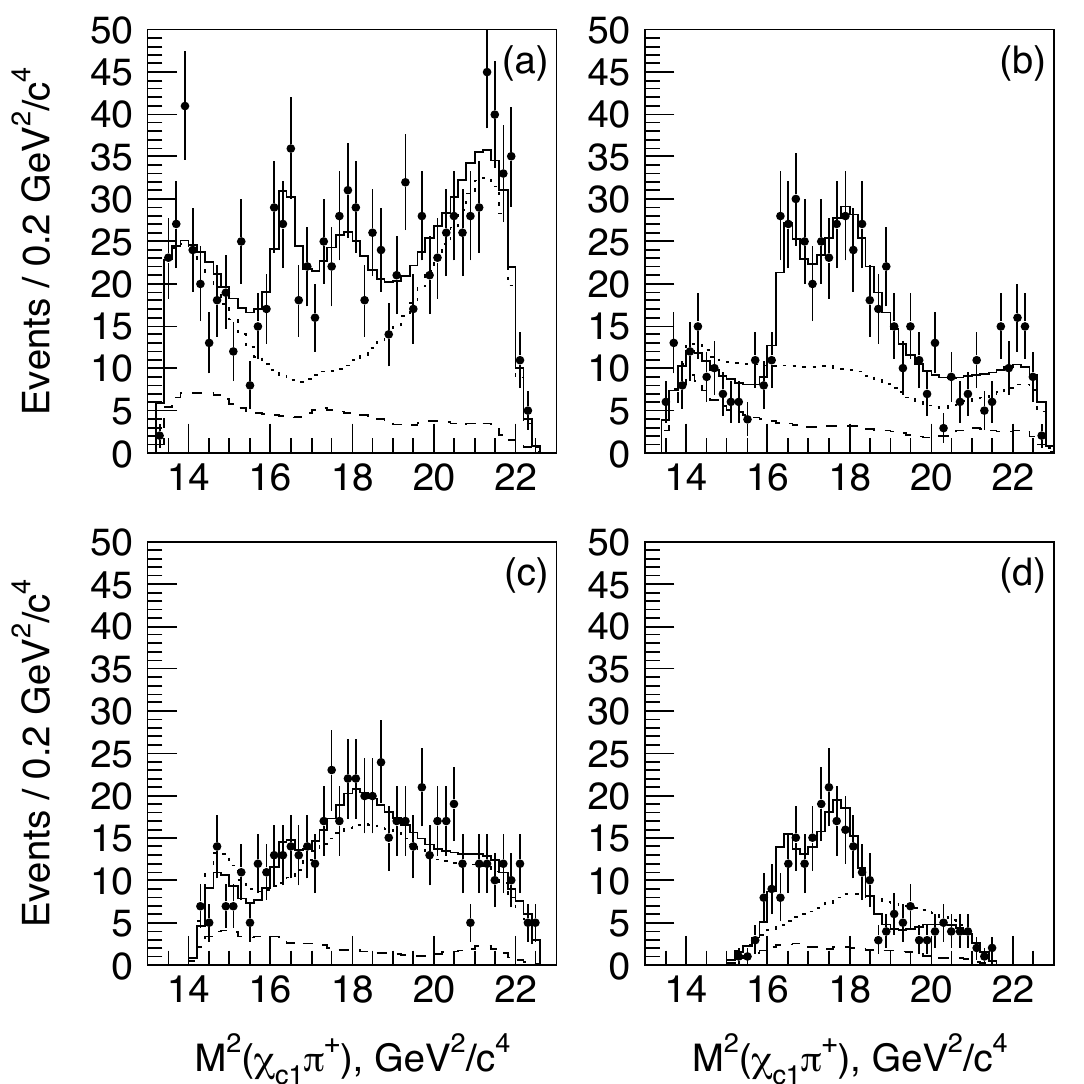}
  \caption{From~\protect\cite{z12-belle}:
	(upper) the $\bzbar \to \kmin \pi^+\chi_{c1}$ Dalitz plot,
	with data- and fit-projections to $M^2(\pi^+\chi_{c1})$ for regions
	(a) $s_x \equiv M^2(\kmin\pi^+) < 1.00$,
	(b) $s_x \in [1.00,1.75]$,
	(c) $s_x \in [1.75,2.37]$, and
	(d) $s_x > 2.37\,\gev^2$,
	fitting with (middle) the known \kstar\ states,
	and (lower) known states plus two interfering 
	$Z^+\to\pi^+\chi_{c1}$ terms.}
  \label{fig-z12-belle-slices}
\end{figure}

The observed $Z(4430)^+$ final state, $\pi^+\psi'$ rather than $\pi^+\psi$,
prompts searches using other charmonia. Belle~\cite{z12-belle}
has studied $\bzbar \to \kmin \pi^+\chi_{c1}$,
finding $2126 \pm 56 \pm 42$ decays
with the Dalitz distribution $(s_x,s_y) = (M^2(\kmin\pi^+),\,M^2(\pi^+\chi_{c1}))$
shown in Fig.~\ref{fig-z12-belle-slices}.
Contributions from $\kbar{}^*(892)^0\chi_{c1}$
and $\kbar{}^*(1430)^0\chi_{c1}$ are visible, 
together with a horizontal band at $M^2(\pi^+\chi_{c1}) \simeq 17\,\gev^2$.

With $\chi_{c1} \to \gamma \psi[\to \ell^+\ell^-]$,
the $\bzbar \to \kmin \pi^+\chi_{c1}$ decay is described by six variables:
$s_x$, $s_y$, 
$\cos\theta_{\chi_{c1}}$,
$\phi_{\chi_{c1}}$,
$\cos\theta_{\psi}$, and
$\phi_{\psi}$.
Belle integrates over the angular quantities;
efficiency is near-uniform in the azimuthal angles $\phi$, while
polar angle distributions are studied as a cross-check. 
A binned likelihood fit to the Dalitz plot is then performed, 
using $S(s_x,s_y) \times \epsilon(s_x,s_y) + B(s_x,s_y)$
where the efficiency $\epsilon$ is determined from Monte Carlo and the 
background distribution $B$ from sidebands; both are smoothed.
The signal function $S$ follows the isobar model,
including all known low-lying $\kmin\pi^+$ resonances
and either 0, 1, or 2 exotic $Z\to\pi^+\chi_{c1}$ terms.
Blatt-Weisskopf form factors,
energy-dependent widths, 
and angular terms (calculated in the helicity formalism)
are used; the latter take into account the different possible $\chi_{c1}$
parents, \bzbar\ and $Z$.

The fit with only the known $\kmes\pi$ resonant states is poor,
failing to resolve an enhancement at $M(\kmin\pi^+)\sim 4150\,\mev$,
and troughs in the distribution
(see Fig.~\ref{fig-z12-belle-slices}, middle plots).
Adding further \kstar, or $\chi_{c1}\kmes$ non-resonant amplitudes,
still leaves a poor fit.
A dramatic improvement is seen with a single $Z^+\to\pi^+\chi_{c1}$ 
term ($>10\sigma$) included, although fine structure is imperfectly
reproduced; the fit probability is 0.1\%.
Adding a \emph{second} $Z^+$ terms leads to a $>5\sigma$ improvement,
and a 40\% fit probability, with good agreement 
between the fit and data throughout
(Fig.~\ref{fig-z12-belle-slices}, lower plots).
A summary of the fits, projected onto one slice of the Dalitz plane,
is shown in Fig.~\ref{fig-z12-belle-display}.

\begin{table}
  \caption{From~\protect\cite{z12-belle}: parameters for the
	exotic $Z^+ \to \pi^+\chi_{c1}$ terms in the Belle
	$\bzbar \to \kmin \pi^+\chi_{c1}$ analysis.}
  \label{table-z12-belle}
  \renewcommand{\arraystretch}{1.3}
  $
  \begin{array}{lll}
    \hline\hline
\multicolumn{1}{l}{~}
& \multicolumn{1}{c}{Z_1^+}	& \multicolumn{1}{c}{Z_2^+}	\\ \hline
M/\mev	& 4051 \pm 14^{+20}_{-41} 			& 4248^{+44}_{-29}{}^{+180}_{-35}		\\
\Gamma/\mev
& \phantom{40}82^{+21}_{-17}{}^{+47}_{-22}	& \phantom{4}177^{+54}_{-39}{}^{+316}_{-61}	\\
(\br_{\bzbar}\times\br_{Z^+}) / 10^{-5}
& (3.1^{+1.5}_{-0.9}{}^{+3.7}_{-1.7})		& (4.0^{+2.3}_{-0.9}{}^{+19.7}_{-0.5})	\\
  \hline\hline
  \end{array}
  $
\end{table}

Thirteen variants of the fits
are performed: a $\pi^+\chi_{c1}$ resonant term is required
at $>6\sigma$ in all cases, 
and two terms are preferred over one at $>5\sigma$.
Parameters of the new candidate states are shown
in Table~\ref{table-z12-belle}.
Product branching fractions
are comparable to those of the $X(3872)$, $Y(3940)$ and $Z(4430)^+$
in \bmes-decays.
In summary, $Z_1^+$ and $Z_2^+\to\pi^+\chi_{c1}$ join the $Z(4430)^+$
as candidates for charged (and hence exotic) hidden-charm states.

\begin{figure}
  \includegraphics[width=5.5cm]{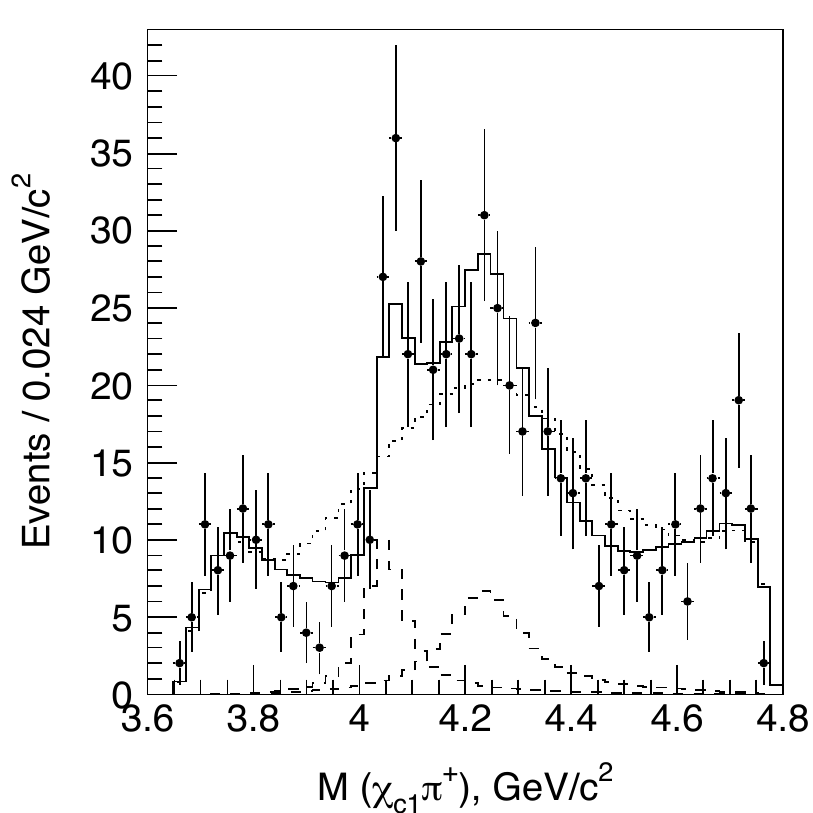}
  \caption{Invariant mass distribution for $\pi^+\chi_{c1}$
	in $\bzbar \to \kmin \pi^+\chi_{c1}$ events
	at Belle~\protect\cite{z12-belle},
	for $M^2(\kmin\pi^+)\in [1.00,1.75]\,\gev$.
	The solid (short-dashed) curve shows the projection
	of the fit with (without) $Z^+\to\pi^+\chi_{c1}\pi^+$ resonant terms;
	contributions of the latter are shown long-dashed.}
  \label{fig-z12-belle-display}
\end{figure}


\section{Appendix: Questions following the talk}
\label{section-questions}

\noindent
\textbf{Q1 (Alan Schwartz):}					\\
How is the $\kmin \pi^+\chi_{c1}$ fit quality determined?	\\
\textbf{A1:} With 92 bins of varying size,
enforcing $\geq 16$ events expectation.
Pearson's $\chi^2$ statistic is used.

\noindent
\textbf{Q2 (A.S.):} 				
Do the $Y(3940)$ parameters agree?	\\
\textbf{A2:} Very poorly, but the fits are very different:
only Belle uses a threshold function for the background
(BaBar uses a Gaussian!), but BaBar fits over a wider range, 
and employs more corrections. Comparison should await comparable
fits. A significant peak at threshold is clear
in both studies.

\noindent
\textbf{Q3 (Ralf Gothe):} What was the explanation of
the $\pi^+\pi^-\psi'$ peak seen in ISR, before Belle split it in two?
How was the width understood?	\\
\textbf{A3:} 
It's agreed that the decay to $\psi'$, rather than $\psi$,
is telling us something---but what?
Eichten notes that the central masses of peaks, for a single pole,
can depend on the final state~\cite{charm2007-panel}.
The ``hadro-charmonium'' idea~\cite{dubynskiy-voloshin} is also interesting.
As to the width: I'm not sure.


\end{document}